\documentstyle[12pt]{article}

\author{\sc
       P.M. Lavrov\thanks{E-mail: lavrov@tspu.edu.ru},\,\,P.Yu. Moshin
       \\
       \normalsize\it
       Tomsk State Pedagogical University, Tomsk 634041, Russia}

\title{\LARGE\bf
       A Superfield Formalism of osp(1,2) Covariant Quantization}

\date{}

\begin{document}

\maketitle

\begin{quotation}
\small
\noindent
We  propose a superfield description of $osp$(1,2) covariant quantization by
extending  the  set  of  admissibility conditions for the quantum action. We
realize  a  superfield  form of the generating equations, specify the vacuum
functional  and  obtain  the  corresponding transformations of extended BRST
symmetry.
\end{quotation}

\section{Introduction}

 It  is  well-known  that gauge field theory provides the universal setting
 for  description  of  the  fundamental  interactions, while the manifestly
 covariant (Lagrangian) quantization of gauge theories in the path-integral
 approach  is  the  most effective formalism for the study of their quantum
 properties.  The  main ingredient of covariant quantization is the concept
 of  generating equations for the quantum action, expressed in terms of the
 corresponding   generating   operators   and   antibrackets   (see,  e.g.,
 \cite{bv,blt,l,glm}).

 The  general  approach  to covariant quantization is based on BRST symmetry
 \cite{brst},  which  is  a  global  supersymmetry  of  the integrand in the
 vacuum  functional.  This symmetry was originally discovered in Yang--Mills
 theories  (quantized  according to the Faddeev--Popov rules) and afterwards
 generalized   to  extended  BRST  symmetry  by  adding  so-called  antiBRST
 transformations  \cite{antibrst}.  Extended BRST symmetry permitted Bonora,
 Pasti  and Tonin \cite{bpt} to discover a superfield description of quantum
 Yang--Mills  theories, where this symmetry is realized as supertranslations
 along additional anticommuting coordinates.

 The  $Sp(2)$  covariant  quantization  \cite{blt}  realized  extended  BRST
 symmetry  for  arbitrary  (general)  gauge  theories,  i.e. theories of any
 stage   of   reducibility   with   a   closed  or  open  algebra  of  gauge
 transformations.  This quantization scheme allows to describe the structure
 \cite{bv}  of  the  complete configuration space $\phi^A$ of a gauge theory
 in  terms  of irreducible representations of the group $Sp(2)$, which leads
 to  considerable simplifications of gauge-fixing compared to the well-known
 BV   formalism   \cite{bv}   based   on   the   standard   BRST   symmetry.

 In  \cite{l},  a superfield form of the $Sp(2)$ covariant scheme \cite{blt}
 was  discovered,  which  provides a superfield description of extended BRST
 symmetry for arbitrary gauge theories. This formalism allows to combine the
 entire  set  of  variables  \cite{blt},  i.e.  the  fields  and  antifields
 $(\phi^A,\phi^*_{Aa},\bar{\phi}_A)$,     the     Lagrangian     multipliers
 $(\pi^{Aa},\lambda^A)$  and the sources $J_A$ for the fields $\phi^A$, into
 superfields   $\Phi^A(\theta)$   and   supersources  $\bar{\Phi}_A(\theta)$
 defined  in a superspace with two anticommuting coordinates $\theta^a$. The
 quantum  action  is  defined  \cite{l}  as  a functional of superfields and
 supersources  which  makes  it possible to represent extended BRST symmetry
 as    supertranslations   (along   the   anticommuting   coordinates)   and
 transformations   generated  by  the  extended  antibrackets,  realized  as
 superfield  modifications  of  the  extended antibrackets introduced in the
 $Sp(2)$ covariant approach \cite{blt}.

 In  \cite{glm},  an  $osp(1,2)$ covariant scheme for general gauge theories
 was  proposed  to  modify the $Sp(2)$ covariant formalism in a way allowing
 to  ensure  the  symplectic invariance of the theory by means of subjecting
 the  quantum  action  to a modified set of generating equations accompanied
 by    analogous   subsidiary   conditions   and   special   conditions   of
 admissibility.  As a consequence, apart form extended BRST symmetry related
 to  the  modified  generating  equations,  the  $osp(1,2)$ covariant scheme
 exhibits   a  new  type  of  global  symmetry  related  to  the  additional
 generating  equations  and  admissibility  conditions. This modification is
 achieved   by  extending  the  original  set  of  variables  \cite{blt}  by
 auxiliary  fields  $\eta_A$  which  allows  to enlarge the set of operators
 encoding  the  generating  equations so that the resulting set of operators
 satisfies  relations isomorphic to the superalgebra $osp(1,2)$ belonging to
 the  class of orthosymplectic superalgebras \cite{alg}. The superalgebra of
 the  generating operators \cite{glm} contains explicit dependence on a mass
 parameter  which  enters  these  operators and is consequently inherited by
 the  quantum  action.  It  is  expected  \cite{glm}  that  the incorporated
 mass-dependence   can   be   applied   to   achieve  an  $Sp(2)$  invariant
 renormalization  of  the  theory.  In  the  massless  limit, the $osp(1,2)$
 covariant   formalism  leads  to  the  standard  $Sp(2)$  covariant  scheme
 considered  in  a  special  case  of  gauge-fixing  and  solutions  to  the
 generating equations.

 In   \cite{gm},   a   superfield   description   of   $osp$(1,2)  covariant
 quantization  was proposed, where the superalgebra $osp(1,2)$ is considered
 as  a  subalgebra  of  the superalgebra $sl(1,2)$, which can be regarded as
 the  algebra of conformal generators in a superspace with two anticommuting
 coordinates.  At  the level of the vacuum functional, the supervariables of
 the  formalism  \cite{gm}  are  identical  to  those applied by the $Sp(2)$
 covariant   superfield  scheme  \cite{glm}  with  the  components  $\eta_A$
 playing  the  role of sources to the fields $\phi^A$.  The aim of the paper
 \cite{gm}  was  to generalize the transformations of extended BRST symmetry
 and  those of the additional global symmetry to the full group of conformal
 transformations, containing the group of translations as a particular case.

 Notice that the superfield description of the extended antibrackets used in
 \cite{gm}   is  different  from  that  applied  by  the  $Sp(2)$  covariant
 superfield  approach  \cite{l}. Namely, the component form of these objects
 is  identical  with  the  extended  antibrackets introduced in the original
 $Sp(2)$  covariant  scheme  \cite{blt}. This choice is due to the fact that
 the  form of extended antibrackets \cite{l} conflicts with the superalgebra
 $sl(1,2)$  satisfied  by  the  symmetry generators \cite{gm}. In fact, this
 situation  takes  place  also in the original formalism \cite{glm} based on
 the superalgebra $osp(1,2)$.

 In  the  recent  paper  \cite{lm}  it  was  demonstrated that the algebraic
 compatibility  of the extended antibrackets with the symmetry generators is
 not  sufficient  for  a  consistent  superfield  quantization. It was shown
 \cite{lm}  that  without  loss  of  generality  the  choice  of  superfield
 antibrackets  in the form \cite{l} is the only one compatible with extended
 BRST  symmetry realized as supertranslations accompanied by transformations
 generated  by  the  extended  antibrackets.  In  particular,  the  form  of
 extended   antibrackets  \cite{blt}  proposed  to  describe  extended  BRST
 symmetry  in  terms  of  conformal  transformations  \cite{gm}  is actually
 incompatible   with   this   symmetry   even  in  the  particular  case  of
 supertranslations.  This  means  that  the problem of consistent $osp(1,2)$
 covariant superfield quantization still remains open.

 To  advance  in  the  solution  of  this  problem,  we  observe,  following
 \cite{lm},  that  the  form  of  extended  antibrackets \cite{blt} could be
 applied  to  a superfield description of extended BRST symmetry in the case
 of  subjecting  the  quantum action to additional restrictions which cancel
 the  non-invariance  related  to  the  particular  choice  of  the extended
 antibrackets.

 In  this  paper  we  propose  a superfield approach to $osp(1,2)$ covariant
 quantization  by means of extending the set of admissibility conditions for
 the  quantum  action  \cite{glm,gm}  on a manifestly superfield basis. This
 allows  to  set  up  a consistent superfield formalism which reproduces the
 original  $osp(1,2)$  covariant  scheme  \cite{glm} as a particular case of
 gauge-fixing.  The  superfield  representation of the extended antibrackets
 used  in  the present formalism was proposed in \cite{gm}. The construction
 of  the  vacuum  functional  is  similar  to  the  approach  \cite{l}  with
 allowance for the necessary modifications.

 The  paper  is  organised  as  follows. In Section 2 we introduce the basic
 definitions  and  notations. In Section 3 we formulate the superfield rules
 of  $osp(1,2)$ covariant quantization. In Section 4 we discuss the relation
 of   the   proposed   formalism   to   the  original  $osp(1,2)$  covariant
 quantization.  In  Section 5 we summarise the results of the paper and make
 concluding remarks.

 We  use  the  conventions adopted in \cite{l,glm}. Derivatives with respect
 to  (super)sources  and  antifields are taken from the left, and those with
 respect  to (super)fields, from the right. Left derivatives with respect to
 (super)fields    are    labelled    by    the    subscript   {\it   ``l''}.

\section{Basic Definitions}

\setcounter{equation}{0}

 Consider  a  superspace ($x^\mu$, $\theta^a$), where $x^\mu$ are space-time
 coordinates,   and  $\theta^a$  is  an  $Sp(2)$  doublet  of  anticommuting
 coordinates.   Notice   that  any  function  $f(\theta)$  has  a  component
 representation,
\[
 f(\theta)=f_0+\theta^a f_a+\theta^2 f_3,
 \;\;\;
 \theta^2\equiv\frac{1}{2}\theta_a\theta^a,
\]
 and an integral representation,
\[
 f(\theta)=\int d^2\theta'\,\delta(\theta'-\theta)f(\theta'),
 \;\;\;
 \delta(\theta'-\theta)=(\theta'-\theta)^2,
\]
 where  raising  and  lowering  the $Sp(2)$ indices is performed by the rule
 $\theta^a=\varepsilon^{ab}\theta_b$,   $\theta_a=\varepsilon_{ab}\theta^b$,
 with    $\varepsilon^{ab}$   being   a   constant   antisymmetric   tensor,
 $\varepsilon^{12}=1$,   and   integration   over  $\theta^a$  is  given  by
\[
 \int d^2\theta=0,\;\;\int d^2\theta\;\theta^a=0,\;\;\int d^2\theta\;
 \theta^a\theta^b=\varepsilon^{ab}.
\]
 In particular, for any function $f(\theta)$ we have
\[
 \int d^2\theta\;\frac{\partial f(\theta)}{\partial \theta^a}=0,
\]
 which implies the property of integration by parts
\begin{equation}
\label{byparts}
 \int d^2\theta\;\frac{\partial f(\theta)}{\partial \theta^a}g(\theta)=
 -\!\int\! d^2 \theta (-1)^{\varepsilon(f)}f(\theta)\frac{\partial g(\theta)}
 {\partial\theta^a},
\end{equation}
 where derivatives with respect to $\theta^a$ are taken from the left.

 We    now    introduce    a    set   of   superfields   $\Phi^A{(\theta)}$,
 $\varepsilon(\Phi^A)=\varepsilon_A$,    with    the    boundary   condition
\[
 \left.\Phi^A(\theta)\right|_{\theta=0}=\phi^A,
\]
 and  a  set  of supersources $\bar{\Phi}_A{(\theta)}$ of the same Grassmann
 parity,     $\varepsilon(\bar{\Phi}_A)=\varepsilon_A$.     The    structure
 \cite{blt}  of  the  complete  configuration  space  $\phi^A$ for a general
 gauge     theory     of     $L$-stage     reducibility    is    given    by
\begin{equation}
\label{ind}
 \phi^A=(A^i,B^{\alpha_s|a_1\cdots a_s},C^{\alpha_s| a_0\cdots a_s}),
 \;\;\;s=0,\ldots,L,
\end{equation}
 where  $A^i$  are  initial classical fields, while $B^{\alpha_s| a_1 \cdots
 a_s}$,  $C^{\alpha_s|  a_0  \cdots  a_s}$  are  pyramids  of  auxiliary and
 (anti)ghost  fields, being completely symmetric $Sp(2)$ tensors of rank $s$
 and $s + 1$, respectively.

 For  arbitrary  functionals  $F=F(\Phi,\bar{\Phi})$, $G=G(\Phi,\bar{\Phi})$
 we define the superbracket operations $(\,\,,\,)^a$, $\{\,\,\,,\,\}_\alpha$
\begin{eqnarray}
 (F,G)^a&=&(-1)^{\varepsilon_A}\int d^2\theta\bigg\{
 \frac{\partial^2}{\partial\theta^2}
 \left(\frac{\delta F}{\delta\Phi^A(\theta)}\right)\theta^a
 \frac{\delta G}{\delta \bar{\Phi}_A(\theta)}
 -\,(F\leftrightarrow G)(-1)^{(\varepsilon (F)+1)
 (\varepsilon (G)+1)}\bigg\},\nonumber\\
\label{bracket}
 \{ F,G \}_\alpha&=&-\!\int\! d^2 \theta\Bigg\{
 \frac{\partial^2}{\partial\theta^2}\left(\frac{\delta F}
 {\delta \Phi^A(\theta)}\right)\theta^2
 \frac{\delta G}{\delta \bar{\Phi}_B(\theta)} (\sigma_\alpha)_B^{~~\!A}
 +\,(F \leftrightarrow G)(-1)^{\varepsilon(F)\varepsilon(G)}\Bigg\},
\end{eqnarray}
 where
\[
 \frac{\partial^2}{\partial\theta^2}\equiv\frac{1}{2}\varepsilon^{ab}
 \frac{\partial}{\partial\theta^b}\frac{\partial}{\partial\theta^a}\;.
\]
 Notice the properties of derivatives
\begin{eqnarray*}
\nonumber
 &&\frac{\delta_l\Phi^A(\theta)}{\delta\Phi^B(\theta^{'})}
 =\delta(\theta^{'}-\theta)\delta^A_B
 =\frac{\delta\Phi^A(\theta)}{\delta\Phi^B(\theta^{'})},\\
\nonumber
 &&\frac{\delta\bar{\Phi}_A(\theta)}{\delta\bar{\Phi}_B(\theta^{'})}
 =\delta(\theta^{'}-\theta)\delta^B_A.
\end{eqnarray*}
 The     elements    of    the    matrix    $(\sigma_\alpha)_A^{~~\!B}\equiv
 -(\sigma_\alpha)^B_{~~\!A}$   in  eqs.~(\ref{bracket})   with  the  indices
 (\ref{ind}) are given by
\begin{equation}
\label{II9}
(\sigma_\alpha)^B_{~~\!A}\equiv\!\left\{
\begin{array}{ll}
 \delta^{\beta_s}_{\alpha_s}(s + 1)(\sigma_\alpha)^b_{~a}
 S^{b_1 \cdots b_s a}_{a_1 \cdots a_s b}
 &A = \alpha_s|a_1 \cdots a_s,\;B = \beta_s|b_1 \cdots b_s,\\
 \delta^{\beta_s}_{\alpha_s}(s + 2)(\sigma_\alpha)^b_{~a}
 S^{b_0 \cdots b_s a}_{a_0 \cdots a_s b}
 &A = \alpha_s|a_0 \cdots a_s,\;B = \beta_s|b_0 \cdots b_s,\\
 0&\mbox{otherwise},
\end{array}
\right.
\end{equation}
 where $S^{b_0 \cdots b_s a}_{a_0 \cdots a_s b}$ is a symmetrizer
 ($X^a$ being independent bosonic variables)
\[
 \!\!\!\!
 S^{b_0 \cdots b_s a}_{a_0 \cdots a_s b} \equiv
 \frac{1}{(s + 2)!} \frac{\partial}{\partial X^{a_0}} \cdots
 \frac{\partial}{\partial X^{a_s}} \frac{\partial}{\partial X^b}
 X^a X^{b_s} \cdots X^{b_0},
\]
 with the properties
\begin{eqnarray*}
 S^{b_0\cdots b_s a}_{a_0 \cdots a_s b}&=&\frac{1}{s + 2}
 \bigg(\sum_{r=0}^s\delta^{b_r}_{a_0}
 S^{b_0 \cdots b_{r-1}b_{r + 1}\cdots b_s a}_{a_1 \cdots a_s b}
 +\frac{1}{s+1}\sum_{r=0}^s\delta^a_{a_0}\delta^{b_r}_b
 S^{b_0\cdots b_{r-1}b_{r+1}\cdots b_s}_{a_1\cdots a_s}\bigg),
 \\
 S^{b_0 \cdots b_s}_{a_0 \cdots a_s}&=&\frac{1}{s+1}
 \sum_{r=0}^s\delta^{b_r}_{a_0}
 S^{b_0\cdots b_{r-1}b_{r+1}\cdots b_s}_{a_1\cdots a_s}.
\end{eqnarray*}
 In eq.~(\ref{II9}) we have
\[
 (\sigma_\alpha)^{ab}=\varepsilon^{ac}(\sigma_\alpha)_c^{~b}
 =(\sigma_\alpha)^a_{~c}\,\varepsilon^{cb}
 =\varepsilon^{ac} (\sigma_\alpha)_{cd}\,\varepsilon^{db},\;\;\;
 (\sigma_\alpha)_a^{~b}=-(\sigma_\alpha)^b_{~a},\;\;\;
 (\sigma_\alpha)^{ab}=(\sigma_\alpha)^{ba},
\]
\begin{equation}
\label{propsigma}
 (\sigma_\alpha)_a^{~a}=(\sigma_\alpha)^a_{~a}=0,\;\;\;
 \varepsilon^{ad} \delta^b_c+\varepsilon^{bd} \delta^a_c
 =-(\sigma^\alpha)^{ab}(\sigma_\alpha)^d_{~c}\,,
\end{equation}
where the matrices $\sigma_\alpha$ ($\alpha = 0,+,-$) form the algebra $sl(2)$
\begin{eqnarray*}
 \sigma_\alpha \sigma_\beta=g_{\alpha\beta}+
 \hbox{$\frac{1}{2}$}\epsilon_{\alpha\beta\gamma}\sigma^\gamma,
 \;\;\;
 \sigma^\alpha=g^{\alpha\beta}\sigma_\beta,
 \;\;\;
 \hbox{Tr}(\sigma_\alpha\sigma_\beta)=2 g_{\alpha\beta},
\end{eqnarray*}
\[
 g^{\alpha\beta} =
 \left(\begin{array}{ccc}
 1 & 0 & 0 \\
 0 & 0 & 2 \\
 0 & 2 & 0
\end{array}\right),
 \;\;\;
 g^{\alpha\gamma} g_{\gamma\beta} = \delta^\alpha_\beta,
\]
 with $\epsilon_{\alpha\beta\gamma}$ being an antisymmetric tensor,
 $\epsilon_{0+-} = 1$.

 Let us introduce a set of first-order operators $V^a_m$, $U^a_m$ (odd)
 and $V_\alpha$, $U_\alpha$ (even),
\begin{eqnarray}
\label{U&V}
 V^a_m&=&\int d^2\theta\bigg\{
 \frac{\partial\bar{\Phi}_A(\theta)}
 {\partial\theta_a}
 \frac{\delta}{\delta\bar{\Phi}_A(\theta)}
 +m^2\bigg((P_{+})^{Ba}_{Ab}\,\theta^b\frac{\partial^2}{\partial\theta^2}
 \left(\theta^2\bar{\Phi}_B(\theta)\right)
 \frac{\delta}{\delta\bar\Phi_A(\theta)}\nonumber\\
 &&-\,\varepsilon^{ab}(P_{-})^{Bc}_{Ab}\,\theta^2
 \frac{\partial^2}{\partial\theta^2}
 \left(\theta_c\bar{\Phi}_B(\theta)\right)
 \frac{\delta}{\delta\bar\Phi_A(\theta)}\bigg)\bigg\},\nonumber\\
 U^a_m&=&\int d^2\theta \bigg\{\frac{\partial{\Phi}^A(\theta)}
 {\partial\theta_a}
 \frac{\delta_l}{\delta{\Phi}^A(\theta)}
 -m^2\bigg((P_{+})^{Ba}_{Ab}\,\theta^2
 \frac{\partial^2}{\partial\theta^2}\left(\theta^b{\Phi}^A(\theta)\right)
 \frac{\delta_l}{\delta\Phi^B(\theta)}\nonumber\\
 &&-\,\varepsilon^{ab}(P_{-})^{Bc}_{Ab}\,\theta_c
 \frac{\partial^2}{\partial\theta^2}
 \left(\theta^2{\Phi}^A(\theta)\right)\frac{\delta_l}{\delta\Phi^B(\theta)}
 \bigg)\bigg\},\nonumber\\
 V_\alpha&=&\int d^2\theta\bigg\{\bar{\Phi}_B(\sigma_\alpha)^B_{~~\!A}
 \frac{\delta}{\delta\bar\Phi_A(\theta)}
 -\frac{\partial^2}{\partial\theta^2}
 \left(\bar{\Phi}_A(\theta)\theta_b\right)
 (\sigma_\alpha)^b_{~a}
 \theta^a\frac{\delta}{\delta\bar{\Phi}_A(\theta)}
 \bigg\}\nonumber,\\
 U_\alpha&=&\int d^2\theta\bigg\{\Phi^A(\sigma_\alpha)_A^{~~\!B}
 \frac{\delta_l}{\delta\Phi^B(\theta)}
 +\frac{\partial^2}{\partial\theta^2}
 \left(\Phi^A(\theta)\theta^a\right)
 (\sigma_\alpha)_a^{~b}
 \theta_b\frac{\delta_l}{\delta\Phi^A(\theta)}\bigg\},
\end{eqnarray}
 where $m$ is a mass parameter. The matrices $(P_{-})^{Ba}_{Ab}$,
 $(P_{+})^{Ba}_{Ab}$ in eqs.~(\ref{U&V}) are given by
\[
 (P_\mp)^{B a}_{A b}=(P_\pm)^{B a}_{A b}-(P_\pm)^B_A \delta^a_b
 +\delta^B_A\delta^a_b,
 \;\;\;
 (P_\pm)^B_A=\delta^b_a (P_\pm)^{B a}_{A b},
 \;\;\;
 (\sigma_\alpha)^B_{~A}=(\sigma_\alpha)^b_{~a}(P_\pm)^{B a}_{A b},
\]
 with
\[
 (P_+)^{B a}_{A b}=
\left\{\begin{array}{ll}
 \delta^i_j \delta^a_b
 &A=i,B=j,
\\
 \delta^{\beta_s}_{\alpha_s} (s + 1)
 S^{b_1\cdots b_s a}_{a_1\cdots a_s b}
 &A=\alpha_s|a_1\cdots a_s,\;B=\beta_s|b_1\cdots b_s,
\\
 \delta^{\beta_s}_{\alpha_s} (s + 2)
 S^{b_0 \cdots b_s a}_{a_0 \cdots a_s b}
 &A=\alpha_s|a_0 \cdots a_s,\;
 B=\beta_s|b_0 \cdots b_s,
\\
 0&\mbox{otherwise.}
\end{array}
\right.
\]
 From the above definitions follow the relations \cite{glm}
\[
 \varepsilon^{ad} (P_\pm)^{B b}_{A d}
 +\varepsilon^{bd} (P_\pm)^{B a}_{A d}
 =-(\sigma^\alpha)^{ab} (\sigma_\alpha)^B_{~A},\\
\]
\[
 \varepsilon^{ad} (P_\pm)^{B b}_{A c} +
 \varepsilon^{bd} (P_\pm)^{B a}_{A c} -
 (\sigma^\alpha)^{ab}(\sigma_\alpha)^e_{~c} (P_\mp)^{B d}_{A e}
 =-(\sigma^\alpha)^{ab}\bigr(
 (\sigma_\alpha)^d_{~c}\delta^B_A +
 \delta^d_c (\sigma_\alpha)^B_{~A}\bigr)
\]
 and the property $(P_\mp)^{A b}_{C d}(P_\pm)^{C d}_{B a} = 0$.

 The operators introduced in eqs.~(\ref{U&V}) obey the $osp(1,2)$
 superalgebra \cite{alg}, with the following non-trivial
 (anti)com\-mu\-ta\-tion relations:
\[
 [V_\alpha,V_\beta]=\epsilon_{\alpha\beta}^{~~~\!\gamma} V_\gamma,
 \;\;\;
 [V_\alpha,V_m^a]=V_m^b(\sigma_\alpha)_b^{~a},
 \;\;\;
 \{V_m^a,V_m^b\}=-m^2(\sigma^\alpha)^{ab} V_\alpha,
\]
\begin{eqnarray}
\label{uvalg}
 &&
 [U_\alpha,U_\beta]=-\epsilon_{\alpha\beta}^{~~~\!\gamma} U_\gamma,
 \;\;\;
 [ U_\alpha, U_m^a ]=-U_m^b (\sigma_\alpha)_b^{~a},
 \;\;\;
 \{U_m^a,U_m^b\}=m^2 (\sigma^\alpha)^{ab} U_\alpha.
\end{eqnarray}
 Let us introduce a set of second-order operators $\Delta^a$ (odd) and
 $\Delta_\alpha$ (even)
\begin{eqnarray}
 \Delta^a&=&\int d^2 \theta\,
 \frac{\partial^2}{\partial\theta^2}\left(
 \frac{\delta_l}{\delta\Phi^A(\theta)}\right)
 \theta^a \frac{\delta}{\delta \bar{\Phi}_A(\theta)}\,,\nonumber\\
\label{DeltaaExSQ2}
 \Delta_\alpha&=&(-1)^{\varepsilon_A+1}\int d^2\theta\,
 \frac{\partial^2}{\partial\theta^2}\left(\frac{\delta_l}
 {\delta \Phi^A(\theta)}\right)\theta^2
 \frac{\delta}{\delta \bar{\Phi}_B(\theta)}\,(\sigma_\alpha)_B^{~~\!A}\,.
\end{eqnarray}
 These operators possess the algebraic properties
\begin{equation}
 \label{deltalg}
 [\Delta_\alpha,\Delta_\beta]=0,
 \;\;\;
 \{\Delta^a,\Delta^b\}=0,
 \;\;\;
 [\Delta_\alpha,\Delta^a]=0,
\end{equation}
\begin{eqnarray}
\label{deltalg2}
 {[}\Delta_\alpha,V_\beta {]}+{[} V_\alpha,\Delta_\beta {]}
 &=&\epsilon_{\alpha\beta}^{~~~\!\gamma}\Delta_\gamma,\nonumber\\
 \{\Delta^a,V_m^b\}+\{V_m^a,\Delta^b\}
 &=&- m^2(\sigma^\alpha)^{ab} \Delta_\alpha,\nonumber\\
 {[}\Delta_\alpha,V_m^a{]}+{[}V_\alpha,\Delta^a{]}
 &=&\Delta^b(\sigma_\alpha)_b^{~a}.
\end{eqnarray}
 From eqs.~(\ref{DeltaaExSQ2}) it follows that the action of the operators
 $\Delta^a$ and $\Delta_\alpha$ on the product of two functionals defines
 the superbracket operations (\ref{bracket}), namely
\begin{eqnarray}
\label{deltabracket}
 \Delta_\alpha (FG) &=& (\Delta_\alpha F) G + F (\Delta_\alpha G) +
 \{ F,G \}_\alpha,\nonumber\\
 \Delta^a (FG) &=& (\Delta^a F) G + F (\Delta^a G) (-1)^{\varepsilon(F)}
 +( F,G )^a (-1)^{\varepsilon(F)}.
\end{eqnarray}
 Using eqs.~(\ref{deltalg}), (\ref{deltalg2}), (\ref{deltabracket}) one can
 derive the properties of the superbrackets at the algebraic level \cite{glm}.

 Let us introduce the operators
\begin{eqnarray}
 \label{Deltas_nExSQ}
 \bar{\Delta}^a_m\equiv\Delta^a+\frac{i}{\hbar}V^a_m,\;\;\;
 \bar{\Delta}_\alpha\equiv\Delta_\alpha+\frac{i}{\hbar}V_\alpha.
\end{eqnarray}
 From eqs.~(\ref{uvalg}), (\ref{deltalg}), (\ref{deltalg2}) it follows that
 these operators obey the superalgebra
\begin{eqnarray*}
\label{mdeltalg}
 {[}\bar{\Delta}_\alpha, \bar{\Delta}_\beta {]} &=& (i/ \hbar)
 \epsilon_{\alpha\beta}^{~~~\!\gamma} \bar{\Delta}_\gamma,
 \nonumber\\
 {[} \bar{\Delta}_\alpha, \bar{\Delta}_m^a{]} &=& (i/ \hbar)
 \bar{\Delta}_m^b (\sigma_\alpha)_b^{~a},
 \nonumber\\
 \{ \bar{\Delta}_m^a, \bar{\Delta}_m^b \}&=& - (i/ \hbar) m^2
 (\sigma^\alpha)^{ab} \bar{\Delta}_\alpha,
\end{eqnarray*}
isomorphic to $osp(1,2)$.

\section{Superfield osp(1,2) Covariant Quantization}
\setcounter{equation}{0}

 Let us consider a superfield analogue of the $osp(1,2)$ covariant
 formalism \cite{glm} constructed along the lines of the $Sp(2)$
 covariant superfield scheme \cite{l}. Define the vacuum functional
 $Z_m$ depending on the mass parameter $m$ as the following path
 integral:
\begin{eqnarray}
\label{ZExSQ}
 Z_m=\int d\Phi\,d\bar{\Phi}\,\exp\bigg\{\frac{i}{\hbar}
 \bigg[W_m(\Phi,\bar{\Phi})-\,\frac{1}{2}\varepsilon_{ab}
 U^a_mU^b_mF(\Phi)+m^2F(\Phi)+\bar{\Phi}\Phi\bigg]\bigg\},
\end{eqnarray}
 where $W_m=W_m(\Phi,\bar{\Phi})$ is the $m$-extended quantum action that
 satisfies the generating equations
\begin{eqnarray}
\label{GEqExSQ}
 \bar{\Delta}^a_m\exp\bigg\{\frac{i}{\hbar}W_m\bigg\}=0,
\end{eqnarray}
 and the subsidiary conditions
\begin{eqnarray}
\label{GEqExSQ2}
 \bar{\Delta}_\alpha\exp\bigg\{\frac{i}{\hbar} W_m\bigg\}=0,
\end{eqnarray}
 with $\bar{\Delta}^a_m$ and $\bar{\Delta}_\alpha$ given by
 eqs.~(\ref{Deltas_nExSQ}). Eqs.~(\ref{GEqExSQ}) and (\ref{GEqExSQ2})
 are equivalent to
\begin{eqnarray}
\label{GEq1ExSQ}
 \frac{1}{2}(W_m,W_m)^a+V_m^a W_m=i\hbar\Delta^a W_m,\\
\label{GEq1ExSQ2}
 \frac{1}{2}\{ W_m, W_m \}_\alpha+V_\alpha W_m=i\hbar\Delta_\alpha W_m,
\end{eqnarray}
 where the superbrackets $(\,\,,\,)^a$, $\{\,\,\,,\,\}_\alpha$ and the operators
 $V_m^a$, $V_\alpha$, $\Delta^a$, $\Delta_\alpha$ are defined by eqs.~(\ref{bracket}),
 (\ref{U&V}), (\ref{DeltaaExSQ2}). The quantum action $W_m$ is assumed to be an
 {\it admissible} solution of eqs.~(\ref{GEq1ExSQ}) and (\ref{GEq1ExSQ2}), which
 implies the fulfillment of the restrictions
\begin{eqnarray}
\label{IV51}
 &&
 \int d^2 \theta \, \theta^2 \left\{
 \frac{\delta W_m}{\delta \bar{\Phi}_A(\theta)}+\Phi^A(\theta)\right\}=0,\\
\label{adm1}
 &&
 \int d^2\theta\,\theta^2\frac{\delta W_m}{\delta\Phi^A(\theta)}=0,\\
\label{adm2}
 &&
 \int d^2\theta\,\theta^a\frac{\delta W_m}{\delta\Phi^A(\theta)}=0.
\end{eqnarray}
 In eq.~(\ref{ZExSQ}), $\bar{\Phi}\Phi$ is a functional of the form
\begin{eqnarray}
\label{BilFExSQ}
 \bar{\Phi}\Phi\equiv\int d^2\theta\,\bar{\Phi}_A(\theta)\Phi^A(\theta),
\end{eqnarray}
 while $F=F(\Phi)$ is a gauge-fixing Boson restricted to the class of $Sp(2)$ scalars
 by the conditions
\begin{equation}
\label{cond2}
 U_\alpha F(\Phi)=0,
\end{equation}
 where $U_\alpha$ are the operators (\ref{U&V}).

 An important property of the integrand in eq.~(\ref{ZExSQ}) is its
 invariance under the following transformations:
\begin{eqnarray}
\label{BRSTExSQ}
 &&
 \delta\Phi^A(\theta)=\mu_a U_m^a \Phi^A(\theta),\;\;\;
 \delta \bar{\Phi}_A(\theta)=\mu_a V_m^a \bar{\Phi}_A(\theta) +
 \mu_a ( W_m, \bar{\Phi}_A(\theta) )^a,\\
 \label{BRSTExSQ2}
 &&
 \delta \Phi^A(\theta)=\mu^\alpha U_\alpha\Phi^A(\theta),
 \;\;\;
 \delta \bar{\Phi}_A(\theta)=\mu^\alpha V_\alpha
 \bar{\Phi}_A(\theta) + \mu^\alpha \{ W_m, \bar{\Phi}_A(\theta) \}_\alpha,
\end{eqnarray}
 where $U^a_m$ are operators given by eqs.~(\ref{U&V}), while $\mu_a$ and
 $\mu^\alpha$ are constant (anti)commuting parameters, $\varepsilon(\mu_a)=1$,
 $\varepsilon(\mu^\alpha)=0$. Eqs.~(\ref{BRSTExSQ})
 realize the transformations of extended BRST symmetry and
 eqs.~(\ref{BRSTExSQ2}) express the symmetry related to the requirement
 of $Sp(2)$ invariance.
 The validity of the symmetry transformations (\ref{BRSTExSQ}),
 (\ref{BRSTExSQ2}) follows from the generating equations (\ref{GEq1ExSQ}),
 (\ref{GEq1ExSQ2}), the admissibility conditions (\ref{adm1}), (\ref{adm2}),
 the conditions (\ref{cond2}) of $Sp(2)$ invariance for the gauge-fixing
 Boson, the algebraic properties (\ref{uvalg}) of the operators $U^a_m$,
 $U_\alpha$, and the properties (\ref{propsigma}) of the matrices
 $\sigma_\alpha$. Besides, it is necessary to use integration by parts
 (\ref{byparts}) as well as to take into account the operator representation
 ($U^a_m$ are first-order operators)
\[
 U^a_m U^b_m F(\Phi) = \{U^a_m,{[}U^b_m, F(\Phi){]}\},\;\;\varepsilon(F)=0
\]
 and the Jacobi identity
\[
 {[}{[}\hat F,\hat G\},\hat H\}
 (-1)^{\varepsilon(\hat F)\varepsilon(\hat H)}
 +{\rm cycl.perm.}\,(\hat F,\hat G,\hat H)\equiv 0
\]
 for the supercommutator ${[}\,\,,\,\}$.

 Notice that the admissibility conditions (\ref{IV51}) have not been used in
 the proof of invariance. These conditions  serve to establish the
 relation of the present superfield formalism to the original $osp(1,2)$
 covariant scheme \cite{glm}. As will be explained in the next section, the
 conditions (\ref{IV51}) are closely related to the absence in the path
 integral (\ref{ZExSQ}) of an additional integration weight used in the
 $Sp(2)$ covariant superfield formalism \cite{l}.

\section{Component Analysis}
\setcounter{equation}{0}

 We now consider the component representation of the formalism
 proposed in the previous section in order to establish its
 relation to the original $osp(1,2)$ covariant scheme \cite{glm}.

 The component form of superfields $\Phi^A(\theta)$ and
 supersources $\bar\Phi_A(\theta)$ reads
\begin{eqnarray*}
 \Phi^A(\theta)&=&\phi^A+\pi^{Aa}\theta_a+\lambda^A\theta^2,\\
 \bar{\Phi}_A(\theta)&=&\bar{\phi}_A-\theta^a\phi^*_{Aa}-\theta^2\eta_A.
\end{eqnarray*}
 The components ($\phi^A, \pi^{Aa}, \lambda^A,\bar{\phi}_A,\phi^*_{Aa},\eta_A$)
 are identical with the set of variables required for the construction of the
 vacuum functional in the $osp(1,2)$ covariant formalism \cite{glm}.

 Denote $F(\Phi,\bar{\Phi})\equiv\tilde{F}(\phi,\pi,\lambda,\bar{\phi},\phi^*,\eta)$.
 The superbrackets $(\,\,,\,)^a$, $\{\,\,\,,\,\}_\alpha$
 in eqs.~(\ref{bracket}) have the following component representation:
\begin{eqnarray}
 (F,G)^a&=&\frac{\delta\tilde{F}}{\delta\phi^A}\;\frac{\delta\tilde{G}}
 {\delta\phi^*_{Aa}}-(\tilde{F}\leftrightarrow\tilde{G})\;
 (-1)^{(\varepsilon(F)+1)(\varepsilon(G)+1)},\nonumber\\
\label{bracket2}
 \{ F,G \}_\alpha&=&(\sigma_\alpha)_B^{~~\!A}
 \frac{\delta\tilde F}{\delta \phi^A} \frac{\delta\tilde G}{\delta \eta_B} +
 (\tilde{F}\leftrightarrow\tilde{G})(-1)^{\varepsilon(F)\varepsilon(G)}.
\end{eqnarray}
 The component form of the second-order operators $\Delta^a$,
 $\Delta_\alpha$ given by eqs.~(\ref{DeltaaExSQ2}) reads
\begin{eqnarray}
 \Delta^a&=&(-1)^{\varepsilon_A}\frac{\delta_{\it l}}{\delta\phi^A}\;
 \frac{\delta}{\delta\phi^*_{Aa}}\,,\nonumber\\
\label{delta2}
 \Delta_\alpha &=& (-1)^{\varepsilon_A} (\sigma_\alpha)_B^{~~\!A}
 \frac{\delta_l}{\delta \phi^A}
 \frac{\delta}{\delta \eta_B}\,.
\end{eqnarray}
 Eqs.~(\ref{bracket2}), (\ref{delta2}) coincide with the superbrackets
 and corresponding delta-operators used in the framework of the $osp(1,2)$
 covariant formalism \cite{glm}.

 The first-order operators $V^a_m$, $V_\alpha$ given by eqs.~(\ref{U&V})
 have the following component representation:
\begin{eqnarray}
 V_m^a&=&\varepsilon^{ab} \phi^*_{A b} \frac{\delta}{\delta \bar{\phi}_A} -
 \eta_A \frac{\delta}{\delta \phi^*_{A a}}+ m^2 (P_+)^{B a}_{A b}
 \bar{\phi}_B \frac{\delta}{\delta \phi^*_{A b}}
 -m^2 \varepsilon^{ab}
 (P_-)^{B c}_{A b} \phi^*_{B c} \frac{\delta}{\delta \eta_A}\,,
 \nonumber\\
 \label{Valpha}V_\alpha &=& \bar{\phi}_B (\sigma_\alpha)^B_{~~\!A}
 \frac{\delta}{\delta \bar{\phi}_A}\, + \bigr( \phi^*_{A b}
 (\sigma_\alpha)^b_{~a} + \phi^*_{B a} (\sigma_\alpha)^B_{~~\!A} \bigr)
 \frac{\delta}{\delta \phi^*_{A a}}
 +\eta_B (\sigma_\alpha)^B_{~~\!A}
 \frac{\delta}{\delta \eta_A}\,.
\end{eqnarray}
 Eqs.~(\ref{bracket2})--(\ref{Valpha}) imply that the superfield
 generating equations (\ref{GEqExSQ}), (\ref{GEqExSQ2}), or
 equivalently (\ref{GEq1ExSQ}), (\ref{GEq1ExSQ2}), formally coincide
 with the generating equations for the quantum action in the framework
 of the $osp(1,2)$ covariant approach \cite{glm}.

 The component form of the first-order operators $U^a_m$, $U_\alpha$ given
 by eqs.~(\ref{U&V}) reads
\begin{eqnarray}
 U_m^a&=&(-1)^{\varepsilon_A}\varepsilon^{ab}
 \lambda^A\frac{\delta_l}{\delta\pi^{Ab}} -(-1)^{\varepsilon_A}\pi^{Aa}
 \frac{\delta_l}{\delta \phi^A}
 +m^2\varepsilon^{ab}(-1)^{\varepsilon_A}(P_-)^{B c}_{A b}\phi^A
 \frac{\delta_l}{\delta\pi^{Bc}}\nonumber\\
 &&
 -\,m^2(-1)^{\varepsilon_A}(P_+)^{B a}_{A b} \pi^{A b}
 \frac{\delta_l}{\delta\lambda^B},\nonumber\\
\label{Ualpha}
 U_\alpha&=&\phi^B(\sigma_\alpha)_B^{~~\!A}
 \frac{\delta_l}{\delta\phi^A}+\bigr(\pi^{A b} (\sigma_\alpha)_b^{~a}
 +\pi^{B a} (\sigma_\alpha)_B^{~~\!A} \bigr)\frac{\delta_l}{\delta\pi^{A a}}
 +\lambda^B (\sigma_\alpha)_B^{~~\!A} \frac{\delta_l}{\delta\lambda^A}\,.
\end{eqnarray}
 The admissibility conditions (\ref{adm1}), (\ref{adm2}) for the quantum action
 $\tilde{W}_m=\tilde{W}_m(\phi,\pi,\lambda,\bar{\phi},\phi^*,\eta)$ can be
 represented as
\begin{eqnarray}
 \label{w1}
 \frac{\delta\tilde{W}_m}{\delta\lambda^A}=
 \frac{\delta\tilde{W}_m}{\delta\pi^{Aa}}=0.
\end{eqnarray}
 These conditions have been introduced in order to compensate
 the non-invariance of the integrand in eq.~(\ref{ZExSQ}) related
 to the specific choice of the superbrackets (\ref{bracket2}).
 At the same time, eqs.~(\ref{w1}) restrict the variables of the
 functional $\tilde{W}_m$ to the set $(\phi^A,\bar\phi_A,\phi^*_{Aa},\eta_A)$,
 which is the complete set of variables entering the quantum action
 in the original $osp(1,2)$ covariant formalism \cite{glm}.
 On the hypersurface (\ref{w1}) the generating equations (\ref{GEq1ExSQ}),
 (\ref{GEq1ExSQ2}) become identical with the generating equations
 of the $osp(1,2)$ covariant scheme.

 The component form of the remaining admissibility condition (\ref{IV51})
\begin{equation}
\label{adm3}
 \frac{\delta\tilde{W}_m}{\delta \eta_A} = \phi^A
\end{equation}
 leads to an additional simplification of the quantum action:
\begin{equation}
\label{simpl}
 \tilde{W}_m={\cal W}_m(\phi,\bar\phi,\phi^*)+\eta_A\phi^A.
\end{equation}
 From the viewpoint of the $osp(1,2)$ covariant approach \cite{glm}, this
 form of dependence on the sources $\eta_A$, with allowance for
 eqs.~(\ref{II9}), (\ref{propsigma}), gives the advantage of transforming
 the generating equations (\ref{GEq1ExSQ2}) into the simplified requirement
 of $Sp(2)$ invariance \cite{glm}
\[
 (\sigma_\alpha)_B^{~~\!A} \frac{\delta\tilde{W}_m}{\delta \phi^A} \phi^B
 +V_\alpha \tilde{W}_m=0.
\]

 Let us restrict the gauge-fixing Boson in the vacuum functional
 (\ref{ZExSQ}) to the class of gauges used in the original $osp(1,2)$
 covariant formalism \cite{glm}, i.e. gauges depending only on the fields:
 $\tilde{F}=\tilde{F}(\phi)$. Then, with allowance for the component
 representation (\ref{Ualpha}) of the operators $U_\alpha$, the condition
 (\ref{cond2}) of $Sp(2)$ invariance imposed on the gauge-fixing Boson
 reduces to
\begin{equation}
\label{sp2}
 (\sigma_\alpha)_B^{~~\!A} \frac{\delta\tilde F}{\delta \phi^A}
 \phi^B = 0,
\end{equation}
 which, according to eq.~(\ref{adm3}), can be represented as
\begin{equation}
\label{equiv}
 (\sigma_\alpha)_B^{~~\!A} \frac{\delta\tilde{F}}{\delta
 \phi^A} \frac{\delta\tilde{W}_m}{\delta \eta_B} = 0.
\end{equation}
 Eqs.~(\ref{sp2}) and (\ref{equiv}) reproduce the whole set of additional
 restrictions imposed on the quantum action and gauge-fixing Boson in the
 $osp(1,2)$ covariant scheme \cite{glm}. Namely, eq.~(\ref{sp2}) is imposed
 to ensure the symplectic invariance of the gauge-fixing Boson \cite{glm},
 while the equation (\ref{equiv}) emerges as a condition of admissibility
 for the quantum action, introduced to provide an $Sp(2)$ invariant
 gauge-fixing \cite{glm}.

 Notice that in the $osp(1,2)$ covariant formalism \cite{glm} the condition
 (\ref{adm3}) is redundant, being in fact a particular solution of a more
 fundamental equation (\ref{equiv}) imposed on the quantum action.
 On the contrary, in the present formalism the status of these two
 conditions is {\it reversed}, namely, eq.~(\ref{equiv}) emerges as
 a consequence of eq.~(\ref{adm3}), corresponding to a particular case
 of gauge-fixing.

 The crucial role of the condition (\ref{adm3}) for the present formalism is
 due to the following reasons. On the one hand, a superfield description of
 the $osp(1,2)$ covariant scheme \cite{glm} in terms of the variables $\Phi^A(\theta)$,
 $\bar{\Phi}_A(\theta)$ requires a {\it linear} dependence of the quantum
 action on the auxiliary fields $\eta_A$. Indeed, if this dependence in the
 original $osp(1,2)$ covariant formalism is more than linear, then the
 vacuum functional must be parameterized by a set of variables \cite{glm}
 which is {\it larger} than the set of components of supervariables. On the
 other hand, the admissibility condition (\ref{adm3}) allows to cancel the
 dependence on $\eta_A$ in the integrand (\ref{ZExSQ}) by observing that the
 functional $\bar\Phi\Phi$ in eq.~(\ref{BilFExSQ}) has the component form
\begin{equation}
\label{phi^2}
 \bar{\Phi}\Phi=\bar{\phi}_A\lambda^A+\phi^*_{Aa}\pi^{Aa}-\eta_A\phi^A.
\end{equation}
 This cancellation does not conflict with the original $osp(1,2)$
 prescription \cite{glm}, where the $\eta$-dependence is integrated out by
 imposing the delta-functional constraint $\delta(\eta)$, being, in fact,
 identical with the integration weight introduced within the $Sp(2)$ covariant
 superfield formalism \cite{l}. At the same time, the presence of this
 integration weight in the case of a non-trivial dependence on $\eta_A$
 violates the invariance of the integrand under the transformations
 (\ref{BRSTExSQ}), (\ref{BRSTExSQ2}) because of the non-invariance of
 $\eta_A$, which can be observed from the component representation
 (\ref{Valpha}) of the symmetry generators $V^a_m$, $V_\alpha$. Naturally,
 in the absence of $\eta$-dependence the integration weight $\delta(\eta)$
 can be omitted, as in the case of the present approach (\ref{ZExSQ}),
 compared to the formalism \cite{l}.

 Concluding, we demonstrate the relation of the vacuum functional
 (\ref{ZExSQ}), given
 in terms of
 $\tilde{W}_m=\tilde{W}_m(\phi,\bar{\phi},\phi^*,\eta)$ and
 $\tilde{F}=\tilde{F}(\phi)$,
 to the vacuum functional of the $osp(1,2)$ covariant approach \cite{glm}.

 Notice that the integration measure in eq.~(\ref{ZExSQ}) has the component
 representation
\begin{eqnarray}
\nonumber
 d\Phi\,d\bar{\Phi}=d\phi\,d\pi\,d\lambda\,d\bar{\phi}\,d\phi^*\,d\eta.
\end{eqnarray}

 Using the component form of the operators $U^a_m$ given by
 eqs.~(\ref{Ualpha}) and integrating out the variables $\eta_A$
 with allowance for eqs.~(\ref{simpl}), (\ref{phi^2}), we represent
 the vacuum functional (\ref{ZExSQ}) in the form
\begin{eqnarray}
 \label{Z1} Z_m=\int d\phi\;d\phi^*\,d\pi\,d\bar{\phi}\,d\lambda\,
 \exp\bigg\{\frac{i}{\hbar}\bigg({\cal W}_m+{\cal X}_m
 +\bar\phi_A\lambda^A+\phi^*_{Aa}\pi^{Aa}\bigg)\bigg\},
\end{eqnarray}
 where the quantum action $\tilde W_m={\cal W}_m+\eta_A\phi^A$
 satisfies eqs.~(\ref{GEq1ExSQ}), (\ref{GEq1ExSQ2}),
 (\ref{adm3}), and the gauge-fixing term ${\cal X}_m$ is given by
\begin{eqnarray*}
 {\cal X}_m=\frac{\delta\tilde F}{\delta \phi^A}\lambda^A
 -\frac{1}{2}m^2(P_-)^A_B \frac{\delta\tilde F}{\delta\phi_B}\phi^A
 -\frac{1}{2}\varepsilon_{ab}\pi^{Aa}\frac{\delta^2\tilde F}
 {\delta \phi^A \delta \phi^B}\pi^{B b} + m^2\tilde F,
\end{eqnarray*}
 with $\tilde F$ subject to eq.~(\ref{sp2}).

 The vacuum functional of the $osp(1,2)$ covariant formalism \cite{glm} can be
 represented as
\begin{equation}
\label{ospvac}
 Z_m=\int d\phi\,{\exp}\{(i/\hbar)S_{m,\,{\rm eff}}\},
\end{equation}
with
\[
 \left.S_{m,\,{\rm eff}}(\phi)=S_{m,\,{\rm ext}}(\phi,\bar{\phi},\phi^*,\eta)
 \right|_{\bar{\phi}=\phi^*=\eta=0},
\]
\[
 {\rm exp}\{(i/\hbar)S_{m,\,{\rm ext}}\}=\hat{U}_m(Y)\,{\rm exp}\{(i/\hbar)S_m\},
\]
 where $S_m=S_m(\phi,\bar{\phi},\phi^*,\eta)$ is the quantum action obeying
 the system of generating equations and subsidiary conditions
 (\ref{GEq1ExSQ}), (\ref{GEq1ExSQ2}), (\ref{adm3}) satisfied
 by $\tilde W_m=\tilde W_m(\phi,\bar{\phi},\phi^*,\eta)$, and $\hat{U}_m(Y)$
 is an operator of the form
\begin{eqnarray*}
 \hat{U}_m(Y)={\rm exp}\bigg\{
 \frac{\delta Y}{\delta \phi^A}\bigg(
 \frac{\delta}{\delta \bar{\phi}_A}
 -\frac{1}{2} m^2 (P_-)^A_B \frac{\delta}{\delta \eta_B}\bigg)
 -\frac{\hbar}{2i}\varepsilon_{ab}
 \frac{\delta}{\delta \phi^*_{A a}}
 \frac{\delta^2 Y}{\delta \phi^A \delta \phi^B}
 \frac{\delta}{\delta \phi^*_{B b}}+\frac{i}{\hbar}m^2Y\bigg\},
\end{eqnarray*}
 where $Y=Y(\phi)$ is a gauge-fixing $Sp(2)$ scalar restricted by the
 conditions (\ref{sp2}) imposed on $\tilde F=\tilde F(\phi)$.

 To establish the identity between the vacuum functionals
 (\ref{Z1}) and (\ref{ospvac}), it is sufficient to set
 $S_m=\tilde{W}_m$, $Y=\tilde{F}$.

\section{Concluding Remarks}

 In this paper we have proposed a superfield description of the
 $osp(1,2)$ covariant quantization formalism \cite{glm}.

 We have found superfield representations of the generating equations
 \cite{glm}, constructed the superfield vacuum functional and found the
 corresponding transformations of extended BRST symmetry as well as the
 transformations of the additional global symmetry related to symplectic
 invariance. We have shown that the component representation of the
 formalism reduces to the original $osp(1,2)$ covariant quantization scheme
 \cite{glm} in a particular case of gauge-fixing.

 On   the  one  hand,  the  present  approach  is  based  on  the  component
 realization  of  extended  antibrackets  in  the form introduced within the
 $Sp(2)$   covariant   scheme   \cite{blt},  which  provides  the  algebraic
 compatibility   of   the  antibrackets  with  the  generators  of  symmetry
 transformations  obeying  the  superalgebra  $osp(1,2)$. On the other hand,
 the  formalism  is  based  on extending the set of admissibility conditions
 \cite{glm,gm}  for  the  quantum  action,  which  allows  to  introduce the
 transformations  of  extended BRST symmetry in a manifestly superfield form
 by  canceling  the  non-invariance  related  to  the specific choice of the
 antibrackets.

 In our opinion, the approach used in this paper to provide a superfield
 description of extended BRST symmetry on the basis of the $osp(1,2)$
 superalgebra of symmetry generators should be considered as an
 intermediate step to the complete solution of the problem of $osp(1,2)$
 covariant superfield quantization. Namely, it should be expected that a
 formalism providing such a solution must contain the $Sp(2)$ covariant
 superfield scheme \cite{l} in the massless limit, which is suggested by the
 relation between the original $Sp(2)$ and $osp(1,2)$ covariant methods.
 The difficulty of the realization of such a program lies in setting up a
 formalism providing compatibility of the extended antibrackets used in the
 $Sp(2)$ covariant superfield scheme with the $osp(1,2)$ superalgebra of
 symmetry generators.

\paragraph{Acknowledgement}

 The work was partially supported by the Russian Foundation for Basic
 Research (RFBR), project 99-02-16617, as well as by the Russian Ministry
 of Education (Fundamental Sciences Grant E00-3.3-461). The work of
 P.M.L. was also supported by INTAS, grant 99-0590, and by the joint
 project of RFBR and Deutsche For\-schungs\-gemeinschaft (DFG), 99-02-04022.

\end{document}